\documentclass[journal,twoside]{IEEEtran}
\usepackage{cite}
\usepackage{amsmath,amssymb,amsfonts}
\usepackage{graphicx}
\usepackage{textcomp}

\usepackage{xcolor}
\usepackage{hyperref}
\usepackage{url}
\usepackage{booktabs}
\usepackage{graphicx}
\usepackage{eurosym}
\usepackage{algorithm}
\usepackage{algpseudocode}
\usepackage[caption=false,font=normalsize,labelfont=sf,textfont=sf]{subfig}

\def\BibTeX{{\rm B\kern-.05em{\sc i\kern-.025em b}\kern-.08em
    T\kern-.1667em\lower.7ex\hbox{E}\kern-.125emX}}
\begin{document}

\title{A Simulation Tool for V2G Enabled Demand Response Based on Model Predictive Control}
\author{Cesar Diaz-Londono,~\IEEEmembership{Member,~IEEE}, 
Stavros Orfanoudakis~\IEEEmembership{Student Member,~IEEE}, 
Pedro~P.~Vergara,~\IEEEmembership{Senior Member,~IEEE}
Peter~Palensky,~\IEEEmembership{Senior Member,~IEEE}, 
Fredy~Ruiz,~\IEEEmembership{Senior Member,~IEEE}, and 
Giambattista~Gruosso,~\IEEEmembership{Senior Member,~IEEE}
\thanks{
Cesar is granted by Politecnico di Milano within the IDEA League Alliance through the IDEA League Fellowship Exchange Program 2023. 
Cesar is funded by FSE-REACT EU, PON ``RICERCA e INNOVAZIONE'' 2014-2020, Azione IV.6 (Green).

Stavros is funded by the HORIZON Europe Drive2X Project 101056934.}
\thanks{Cesar Diaz-Londono, Fredy Ruiz, Giambattista Gruosso are with the Dipartimento di Elettronica, Informazione e Bioingegneria, Politecnico di Milano, Milan, Italy (email: \{cesar.diaz, fredy.ruiz, giambattista.gruosso \}@polimi.it)}
\thanks{
Stavros Orfanoudakis, Peter Palensky, and Pedro P. Vergara are with the Intelligent Electrical Power Grids (IEPG) Section, Delft University of Technology, Delft, The Netherlands (emails: \{s.orfanoudakis, p.palensky, p.p.vergarabarrios\}@tudelft.nl).
}

}%
\maketitle

\begin{abstract}
Integrating electric vehicles (EVs) into the power grid can revolutionize energy management strategies, offering both challenges and opportunities for creating a more sustainable and resilient grid.
In this context, model predictive control (MPC) emerges as a powerful tool for addressing the complexities of Grid-to-vehicle (G2V) and vehicle-to-grid (V2G) enabled demand response management.
By leveraging advanced optimization techniques, MPC algorithms can anticipate future grid conditions and dynamically adjust EV charging and discharging schedules to balance supply and demand while minimizing operational costs and maximizing flexibility.
However, no standard tools exist to evaluate novel energy management strategies based on MPC approaches. Our research focuses on harnessing the potential of MPC in G2V and V2G applications, by providing a simulation tool that allows to maximize EV flexibility and support demand response initiatives while mitigating the impact on EV battery health.
In this paper, we propose an open-source MPC controller for G2V and V2G-enabled demand response management. The proposed approach is capable of tackling the uncertainties inherent in demand response operations. Through extensive simulation and analysis, we demonstrate the efficacy of our approach in maximizing the benefits of G2V and V2G while assessing the impact on the longevity and reliability of EV batteries. Specifically, our controller enables Charge Point Operators (CPOs) to optimize EV charging and discharging schedules in real-time, taking into account fluctuating energy prices, grid constraints, and EV user preferences.
\end{abstract}

\begin{IEEEkeywords}
Electric Vehicles (EVs), Smart Charging,
Open Source,
Battery Degradation,
Vehicle-to-Grid (V2G),
Model Predictive Control (MPC)
\end{IEEEkeywords}

\section{Introduction}


\IEEEPARstart{A}{s} the adoption of electric vehicles (EVs) continues to surge, the efficient management of charging infrastructure becomes increasingly important\cite{Tan2016IntegrationOE}.
Grid-to-vehicle (G2V) technology facilitates energy flow in a unidirectional manner, whereas vehicle-to-grid (V2G) technology enables bidirectional energy flow between EVs and the grid, thereby unlocking the potential for EV batteries to serve as flexible energy storage resources~\cite{numberofEVs, Shi2019}. However, the effective implementation of G2V and V2G strategies faces numerous uncertainties, including fluctuating energy demands, variable renewable energy generation, inflexible loads, and dynamic EV schedules~\cite{ALLEHYANI2024247}.

Several mathematical programming methods have emerged for tackling the smart charging problem. For instance, in~\cite{Shamsdin2019Plug-inLot}, a real-time approach is introduced to manage EV charging schedules using satisfaction assignment and prioritized charging strategies, all while considering operational constraints.
Additionally, in~\cite{Schulte2021LinearVehicles}, a linear model is employed to maximize the utilization of PV power generation during EV charging peaks, while ensuring the EVs' state of charge (SoC) constraints are met. Furthermore, Linear Programming (LP) techniques have been utilized to optimize load factor during daily operations of EV parking lots, considering uncertainties in EV behavior such as arrival and departure times~\cite{Sengor2019OptimalUncertainty}. Similarly,~\cite{Meenakumar2020OptimalCapabilities} proposes a mixed-integer programming model to maximize revenues for a V2G EV aggregator in the UK electricity market. In~\cite{Giordano2023}, the synchronization of forecast-based optimization with management operation is implemented for EV charging and discharging patterns. Optimal dispatching methods can reduce the curtailment of wind turbines and PV power~\cite{Jiang2023} or reduce the generation cost~\cite{Patil2021} in distribution systems. While these approaches provide linear models for both small and large-scale problems, they are limited due to the absence of real-time control to handle the inherent uncertainties.

Model Predictive Control (MPC) stands out as a powerful solution for navigating the intricacies of G2V and V2G-enabled demand response management~\cite{9994625}. In~\cite{DIAZLONDONO2024}, a real-time MPC strategy was proposed, effectively reducing operational costs by treating EVs as flexible loads, thereby enabling adjustments in the power allocation and charging duration for each EV. Similarly, in~\cite{TAHMASEBI2021106931}, an MPC method was introduced to coordinate EVs on a large scale, enhancing grid robustness and cutting operational expenses. Moreover, \cite{en9110973} proposed a scenario-based MPC approach, combining robust and stochastic models to minimize the total operational cost for energy management.
In~\cite{ZHOU2021113721}, a real-time cost-minimization energy management strategy was introduced, aiming to maximize the economic potential of EVs while addressing battery degradation, thus mitigating a single vehicle's operating costs.
Even though many MPC methods have been developed, there remains a gap in achieving a comprehensive real-time demand response solution while assessing battery degradation.

Furthermore, while there are many open-source simulation tools for fleet charging management, they often lack baseline MPC implementations. For instance, FleetRL~\cite{Cording_Thakur_2024} provides a customizable environment for commercial vehicle fleets, utilizing Reinforcement Learning (RL) for charging optimization, assessing factors like economic feasibility, battery health, and operational efficiency. Similarly, Chargym~\cite{karatzinis_chargym_2022} and SustainGym~\cite{yeh_sustaingym_2023} offer simulation environments for EV charging, focusing on RL algorithm development but with simplified models. Meanwhile, OPEN~\cite{MORSTYN2020115397} addresses broader smart energy systems but lacks MPC integration. Considering the effectiveness of MPC in EV smart charging, a comparative analysis between existing implementations and those incorporating MPC would be invaluable.

To bridge the identified gaps, we introduce open-source\footnote{Access the open-source code at \url{https://github.com/CesarDiazLondono/MPC-G2V-V2G} (will be public upon paper acceptance).} MPC algorithms embedded within the EV2Gym~\cite{orfanoudakis2024ev2gym} platform, aimed at maximizing profits for Charge Point Operators (CPOs), while accommodating the preferences of EV owners and addressing uncertainties inherent in EV smart charging problems, such as inflexible loads, PV generation, and demand response events. In summary, our primary contributions are:
\begin{itemize}
    \item Development of a flexible open tool capable of executing various EV CPO case studies to evaluate different performance indicators.
    \item Novel formulation of an economic MPC method tailored for G2V and V2G real-time smart charging, strategically crafted to optimize profits while accommodating uncertainties intrinsic to demand response.
    \item Innovative design of the Optimal control with maximum flexibility (OCMF) MPC method for G2V and V2G real-time smart charging, geared towards enhancing flexibility and profitability amidst the complexities of demand response uncertainties.
    \item Evaluation of the impact of smart charging strategies on the degradation of EV batteries.
\end{itemize}

The paper proceeds as follows: We start by providing background information on the simulation platform used, followed by a detailed introduction to smart charging, including mathematical modeling formulations. Finally, we present a thorough experimental evaluation to assess the effectiveness of the proposed strategies.






\section{The EV2Gym Simulator Platform}
\label{sec:ev2gym}

EV2Gym~\cite{orfanoudakis2024ev2gym} provides a complete simulator platform for conducting realistic V2G simulations, encompassing configurable simulation parameters, realistic EV models, charging station management, and complete transformer operations. 

\begin{figure}[t]
    \centering
    \includegraphics[width=1\linewidth]{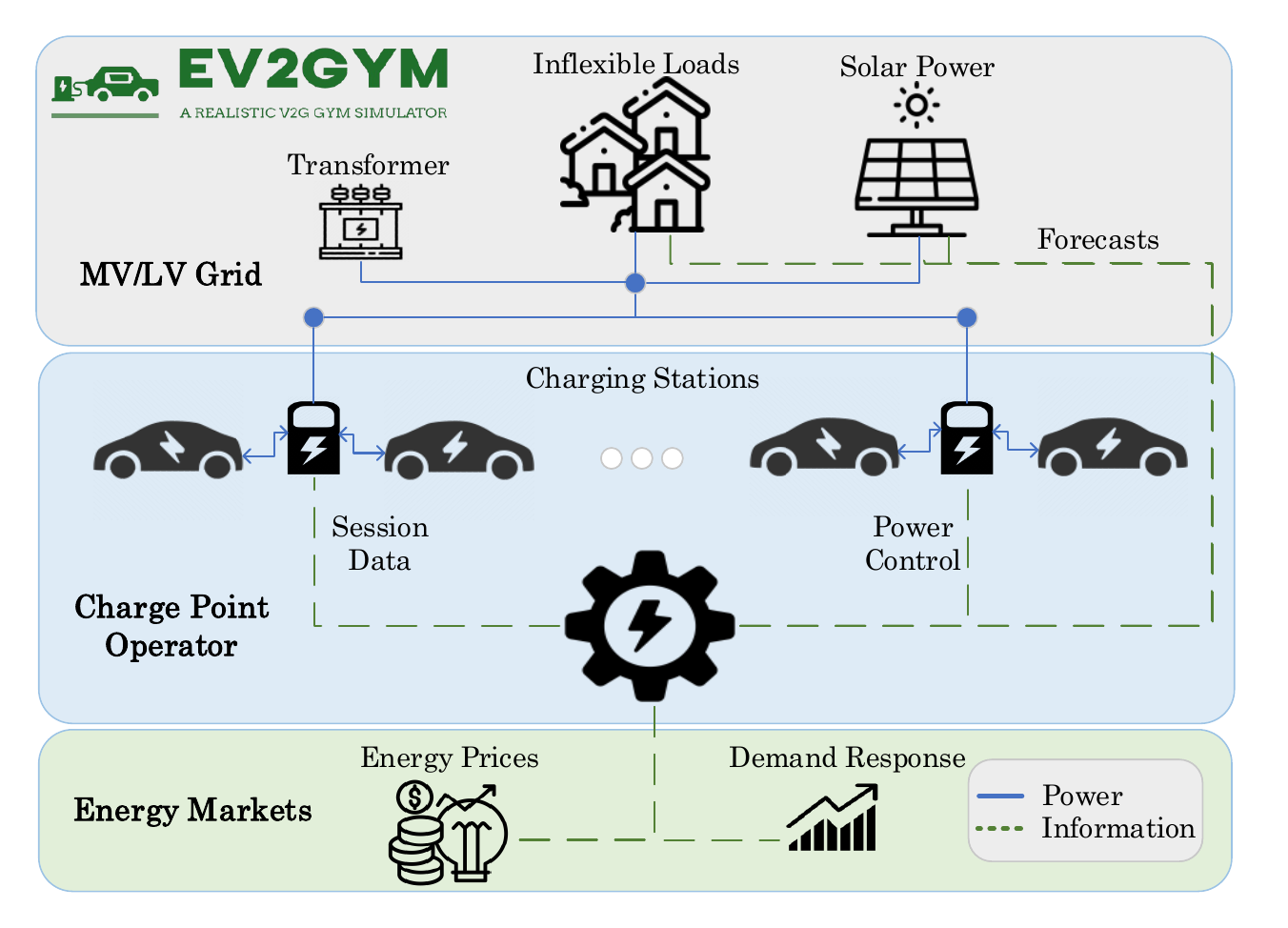}
    \vspace{-8mm}
    \caption{Overview of EV2Gym's modules power and information flow.}
    \label{fig:simulator}
\end{figure}

\subsection{Simulation Models}

In a simulation, there are $I$ chargers, where $i = 1,2,\dots,I$, each one having one Electric Vehicle Supply Equipment (EVSE). Each charger $i$ is connected to one of the $G$ transformers, where $g=1,2,\dots,G$. There are also, $J$ EVs that connect to EVSEs during the simulation. Finally, a simulation consists of $K$ discrete time steps, where $k= 1,2,\dots,K$.
At each step, the user-selected algorithm defines the power output $P_{i,k}$, for each charger $i$. 
The algorithm can fully observe the current state of the environment and can access future data, including electricity prices, load and PV forecasts, and EV schedules. Fig.~\ref{fig:simulator} illustrates the simulator models.

Each EV $j$ in the simulator features customizable parameters, including battery capacity ($E_j$), charge ($\eta^{\textit{c}}$) and discharge ($\eta^{\textit{d}}$) efficiency, minimum SoC at departure $ \underline{SoC}_{j,d_j}$, and maximum charging ($\overline{P}^c_i$) and discharging ($\overline{P}^d_i$) power limits that depend not only on the EV battery management system but also on the EVSE rated power.  
At each time step $k$ EVs can arrive at a charging station based on realistic EV behavior probability distributions obtained from the ElaadNL repository~\cite{elaad}, ensuring the simulation accurately reflects real-world scenarios. 
Therefore, an EV's arrival time $a_j$, departure time $d_j$, and SoC at arrival $SoC_{a_j}$ are sampled from ElaadNL data.
By leveraging this data, EV2Gym can replicate various charging patterns observed in different settings, such as public charging stations, workplaces, and residential areas.

Each transformer $g$ aggregates EV chargers in the set $\Phi_g$, to simulate the electricity distribution network and its constraints. Additionally, transformers incorporate realistic data about inflexible loads, solar photovoltaic (PV) generation, and demand response events. 
For instance, datasets from Pecan Street~\cite{pecanstreet} and Renewables.ninja~\cite{PFENNINGER20161251} are used to model inflexible loads and PV generation profiles, respectively, while information about demand response events can be configured by users based on specific parameters such as start time and duration.
Transformers have operational constraints that must be followed. Violations of these constraints are tracked throughout the simulation, enabling the evaluation of charging strategies under realistic operating conditions. 
Furthermore, the platform generates forecasts of inflexible loads and solar power generation to support the development of algorithms that can handle uncertainty, enhancing the robustness and reliability of charging strategies in real-world scenarios.

\subsection{Battery Degradation Model}

\begin{table}[!t]
\centering
\caption{Battery Degradation Model Parameters}
\label{tab:batt_degrad_constants}
\resizebox{\columnwidth}{!}{%
\begin{tabular}{@{}cccccccc@{}}
\toprule
$\epsilon_0$ & $\epsilon_1$ & $\epsilon_2$ & $\theta$ & $\zeta_0$ & $\zeta_1$ & 
$T^\textit{tot}$ & $Q^{acc}$ \\ \midrule
$6.23\cdot10^6$ & $1.38\cdot10^6$ & $6976$ & $28$ & $4.02\cdot10^{-4}$ & 
$2.04\cdot10^{-3}$ & $730$ & $11160$ \\ \bottomrule
\end{tabular}%
}
\end{table}

Users often hesitate to offer their EVs for V2G services due to concerns about battery degradation. For this reason, it is important to assess the impact of the proposed smart charging algorithms
using a validated battery degradation model~\cite{LEE20236624}. This model comprises both calendar ($d^{\textit{cal}}$) and cyclic ($d^\textit{cyc}$) capacity loss components. For a single EV, the capacity degradation attributed to calendar aging over a simulation with duration $T$ is related to the mean $\widehat{\textit{SoC}}$ and is described as:
\begin{equation}
     d^\textit{cal} = 0.75 \cdot(\epsilon_0 \cdot \widehat{\textit{SoC}} - \epsilon_1)\cdot \text{exp} \left(-\frac{\epsilon_2}{\theta} \right) \cdot \frac{T}{(T^\textit{tot})^{0.25}}.
\end{equation}

Here, $T^\textit{tot}$ denotes the battery's age in days, $\theta$ represents the battery temperature ($^{\circ}$C), and $\epsilon_{0},\epsilon_{1},\epsilon_{2}$ are constants, as shown in Table~\ref{tab:batt_degrad_constants}. 

The cyclic capacity loss depends on the total energy exchanged by the battery and the SoC at each simulation step as:
\begin{equation}
    d^\textit{cyc} = \left( \zeta_0 + \zeta_1 \frac{\sum |\widehat{\textit{SoC}}- \textit{SoC}_k|\,\Delta t}{T}\right) \cdot \frac{\sum |P_k|\, \Delta t}{\sqrt{Q^\textit{acc}}},
    \label{eq:dcyc}
\end{equation}
where $Q^\textit{acc}$ is the accumulated throughput during the battery's lifetime, and $\zeta_0$, $\zeta_1$ are constants defined in Table~\ref{tab:batt_degrad_constants}. Therefore, the total capacity loss of an EV battery is the sum: $Q^\textit{lost}~=~d^\textit{cal} + d^\textit{cyc}$. In this work, the capacity loss is evaluated but not included in the vehicle model within the charging strategies presented in the next section.

\section{Smart Charging Strategies}

The smart charging strategies adopted by the EV Charging Point Operator (EV CPO) rely on the model predictive control (MPC) algorithm. The proposed strategies are based on~\cite{DiazL2019}, in which only unidirectional strategies are considered. In this paper, we are proposing four distinct smart charging algorithms, each designed to achieve specific objectives while adhering to the MPC framework. These strategies include:
\begin{itemize}
\item eMPC G2V: Unidirectional economic MPC (eMPC) aimed at minimizing the operational costs of the CPO.
\item eMPC V2G: Bidirectional eMPC aimed at minimizing the operational costs of the CPO.
\item OCMF G2V: Unidirectional Optimal Control with Maximum Flexibility (OCMF) aimed at minimizing the CPO's operational costs while maximizing charger flexibility.
\item OCMF V2G: Bidirectional OCMF aimed at minimizing the CPO's operational costs while maximizing charger flexibility.
\end{itemize}

\subsection{Solution Formulation}

The proposed MPC algorithms are designed to maximize CPO profits, taking into consideration constrains on energy and power limits imposed by the grid, chargers and vehicles, the preferences of EV owners and the various uncertainties associated with the charging pool, such as the inflexible loads and the PV generation.
MPC algorithm forms the basis of this approach, where at each time step $k$, an optimization program is solved to determine the optimal power exchange between each charger in the pool $I$ with an EV connected, and the grid for an horizon of $H$ steps. Then, the algorithm outputs are the demanded and delivered charger power levels for the current time step $k$, denoted as $P^{c*}_{i,k}$ and $P^{d*}_{i,k}$, respectively. Notice that only one of these power levels can be activated for every charger at each interval $k$, i.e., the charger can only charge or discharge the EV at $k$. On the other hand, the plant under control is the set of EVs connected to the chargers. Power commands $P^{c*}_{i,k}$ and $P^{d*}_{i,k}$ are the inputs to the actual EV charging pool, where the State-of-Charge (SoC) of the vehicles' batteries, represented as $x_{i,k}$, are the system state and model outputs.

The MPC methodology involves four main steps:
\begin{enumerate}
    \item Given the SoC of the connected vehicles at time $k$, build a prediction model to compute the evolution of the SoC of the EV charging pool $x_{i,k+h}$ over a prediction horizon $H$ for any power profiles $P^{c}_{i,k+h}$ and $P^{d}_{i,k+h}$, where $h~=~1,2,\dots, H$.
    \item Optimization process to determine the future inputs $P^{c*}_{i,k+h}$ and $P^{d*}_{i,k+h}$, considering a cost function and operational constraints.
    \item Application of the first element of the optimal power profile $P^{c*}_{i,k}$ or $P^{d*}_{i,k}$ ($h{=}0$) to the chargers within the pool.
    \item Wait until the next time stem to measure the new system state $x_{i,k+1}$ and forecasted variables, such as the inflexible loads and the PV generation.
\end{enumerate}

These steps are iterated at each time step $k$, employing the concept of receding horizon. This approach involves recalculating the power signals $P^{c*}_{i,k}$ and $P^{d*}_{i,k}$ at each iteration, taking into account updated information.

The MPC algorithm relies on a system model to predict future powers in the charging pool. Therefore, this model must strike a balance between precision, to factor in charger responses, and simplicity, to facilitate online optimization for the Real-Time (RT) operation of the chargers. Hence, the EV charger model depends on a signal $\xi_{i,k}$ indicating whether an EV is plugged in at the $k$-th time slot, along with the SoC variation of the plugged-in EV. The signal $\xi_{i,k} $ is a binary variable, $\xi_{i,k} =1$ if the charger $i$ has a plugged-in EV.

The SoC evolution on the $j$-th EV plugged-in the charger $i$ (i.e., $\xi_{i,k} =1$) is:
\begin{equation}
\label{eq:EVch}
    x_{i,k+1} = x_{i,k} + \frac{\Delta t}{E_j} \left ( \eta^{c} P^c_{i,k} - \frac{P^d_{i,k}}{\eta^{d}} \right ),
\end{equation}
while the state space model for an EV charging pool considers three possible conditions, as:
\begin{equation}
	\label{eq:EVSE}
	x_{i,k+1} = \left\lbrace
	\begin{array}{ll}
        \widetilde{SoC}_{j,a_j}  & \textup{if } \quad k = a_j, \\
	\text{Eq.~} \eqref{eq:EVch} & \textup{if } \quad  a_j < k < d_j, \\
	0            & \textup{if } \quad \xi_{i,k} {=} 0 \lor  k {=} d_j.  \\
	\end{array}
	\right.
 \end{equation}

The first condition corresponds to the estimated arrival SoC, denoted as $\widetilde{SoC}_{j,a_j}$. The second condition represents the evolution of the energy stored in the EV battery. The third condition applies when there is no EV connected or when an EV is departing.

Then, the EV charging pool model for the time slot $k$ is defined as:
\begin{equation}
\label{eq:PoolModelk}
    \hat{\mathrm{x}}_{k+1} = A_{k} \hat{\mathrm{x}}_{k} + B_{k} \hat{\mathrm{P}}_{k},
\end{equation}
where, the charging pool outputs, i.e., the SoC in the chargers are presented in the $\hat{\mathrm{x}}_{k} \in{\rm I\!R}^{I} $, as:
\begin{equation}
    \hat{\mathrm{x}}_{k} = [x_{1,k} \ x_{2,k} \ \dots \ x_{I,k}]^\text{T}, \nonumber
\end{equation}
and the exchanged power pool inputs, i.e., the charging and discharging powers are grouped in one vector denoted $\hat{\mathrm{P}}_{k}\in~{\rm I\!R}^{2I}$, as:
\begin{equation}
   \hat{\mathrm{P}}_{k} = [P^c_{1,k} \ P^d_{1,k} \ P^c_{2,k} \ P^d_{2,k} \ \dots \ P^c_{I,k} \ P^d_{I,k}]^\text{T}. \nonumber
\end{equation}

Moreover, $A_{k} = \text{diag}(\xi_{i,k})$ containing information about the plugged-in EVs in the chargers, then, $A_{k} \in~{\rm I\!R}^{I \times I} $. Considering $b^c_{i,k}$ and $b^d_{i,k}$ as:
\begin{equation}
    b^c_{i,k} = \frac{\xi_{i,k} \Delta t \eta^c}{E_j} \quad \land  \quad b^d_{i,k} = \frac{\xi_{i,k} \Delta t}{E_j \eta^d},
\end{equation}
the matrix $B_{k} \in~{\rm I\!R}^{I \times 2I}$ is generated as:
\begin{equation}
\footnotesize
    B_{k} {=} 
    \begin{bmatrix}
    b^c_{1,k} & -b^d_{1,k} & 0  & 0 & \cdots & \cdots & 0 & 0 \\
    0 & 0  & b^c_{2,k} & -b^d_{2,k} & 0 & 0 & 0 & 0 \\
    \vdots  & \vdots& 0 & 0 & \ddots & \ddots & 0 & 0 \\
    0 & 0 & 0 & 0 & \cdots & \cdots & b^c_{I,k} & -b^d_{I,k} \end{bmatrix}.
\end{equation}

Note that the model in Eq.~\eqref{eq:PoolModelk} accounts for all chargers during time slot $k$. For the prediction horizon $H$, the MPC algorithm utilizes the following prediction model.
\begin{equation}
	\label{eq:EvoX}
	\mathbb{X} = \mathbb{A} \hat{\mathrm{x}}_{k} + \mathbb{G} \mathbb{P},
	\end{equation}
where,
\begin{equation}
        \scriptsize
	\begin{split}
	\label{eq:EvoX2}
	\mathbb{X} {=} 
	\begin{bmatrix}
    \hat{\mathrm{x}}_{k+1} \\ \hat{\mathrm{x}}_{k+2} \\ \vdots \\ \hat{\mathrm{x}}_{k+H} \end{bmatrix}, \quad 
	\mathbb{P} = \begin{bmatrix} \hat{\mathrm{P}}_{k} \\ \hat{\mathrm{P}}_{k+1} \\ \vdots \\ \hat{\mathrm{P}}_{k+H-1} \end{bmatrix}, \quad
	\mathbb{A} = \begin{bmatrix}
    A_{k} \\ A_{k} {A}_{k+1} \\ \vdots  \\ {A}_{k} \cdots {A}_{k+H-1} \\     
	\end{bmatrix}, \\
 	\mathbb{G} {=} \begin{bmatrix}
    {B}_{k}               & 0             & 0      & 0 \\
    {A}_{k+1} {B}_{k} & {B}_{k+1} & 0      & 0 \\
    \vdots                      & \vdots        & \ddots & 0 \\
    {A}_{k+1} \cdots {A}_{k+H-1} {B}_{k} & {A}_{k+2} \cdots {A}_{k+H-1} {B}_{k+1} 	& \cdots & {B}_{k+H-1} \end{bmatrix}
	\end{split}
\end{equation}

The algorithm considers the constraint that the SoC at the departure time $d_j$ of each EV must be within predetermined bounds, i.e., the minimum $\underline{SoC}_{j,d_j}$ and maximum $\overline{SoC}_{j,d_j}$ SoC levels specified by the EV owner.
\begin{equation}
    \label{eq:ConsSoC}
    \underline{SoC}_{j,d_j} \leq x_{i,d_j} \leq \overline{SoC}_{j,d_j} 
\end{equation}

The SoC in the charger is always positive and depends on the maximum capacity $\overline{x}_{i}$ of the plugged-in EV battery, as:
\begin{equation}
    \label{eq:SoC}
    \underline{x}_{i, k} \leq x_{i,k} \leq \overline{x}_{i}.
\end{equation}

To avoid deep discharges in the EV battery, a minimum SoC level is defined:
\begin{equation}
\small
    \label{eq:SoCmin}
    \underline{x}_{i, k} {=} \left\lbrace
    \begin{array}{ll}
        \widetilde{SoC}_{j,{a_j}}  & \textup{if } \ k = a_j, \\
        x_{i,k-1} & \textup{if } \  \xi_{i,k} {=} 1 \land x_{i,k-1} < SoC_{j,min},\\
        SoC_{j,min}  & \textup{if } \ \xi_{i,k} {=} 1  \land x_{i,k-1} > SoC_{j,min}, \\
    0            & \textup{if } \ \xi_{i,k} {=} 0 .  \\
	\end{array}
	\right.
 \end{equation}

The first condition for the minimum SoC assumes the arrival SoC at the time of arrival. The second and third conditions depend on the minimum state of charge, $SoC_{j,min}$, which represents the threshold at which an EV is allowed to discharge. Consequently, if the actual SoC of the EV is below this threshold, the minimum SoC remains at the actual SoC level until the threshold is reached. Once the threshold is exceeded, the EV battery can be discharged up to the threshold. The final condition applies when no EV is plugged in.

Moreover, commercial EVSE units are limited to a predefined range of current pilot signals~\cite{SAEJ1772}. As per this specification, during EV charging, the lowest pilot current allowed is 6~A, while the highest depends on the charger's capacity, with discrete control occurring in increments of 1~A. The voltage in the charger is assumed to be constant. Consequently, the power levels are as follows,
\begin{equation}
    \label{eq:pilotSig}
        P^c_{i,k} = P^d_{i,k} = V I_{i,k}, \ \ \ \   I_{i,k} \in \{ 0, 6,7, \dots, 32 \}.
\end{equation}

The algorithm is capable of reacting to RT market signals, including fluctuations in price and demand response (DR) events. Specifically, DR requests are represented by the variable $P^{DR}_k$, allowing the algorithm to promptly respond to market requests as they occur.

The power limit imposed by the each transformer is specified as $\Psi_g$, then the balance equation for transformer $g$ is constrained as
\begin{equation}
    \label{eq:ConsTra}
    \sum \phi_{g,k} + \Tilde{l}_{g,k} - \Tilde{w}_{g,k} \leq  \Psi_{g} - P^{DR}_{g,k}
\end{equation}
where the net power consumed/supplied by the chargers connected to transformer $g$ is $\phi_{g,k} = {P^c_{i,k} - P^d_{i,k}, \forall i \in \Phi_g }$, and $\Phi_g$ denotes the set of EV chargers connected to cluster $g$. 
 $\Tilde{l}_{g,k}$ is the predicted inflexible load, $\Tilde{w}_{g,k}$ is the predicted PV generation and $P^{DR}_{g,k}$ is the demand response command is case of a market request.


In the following subsections, the MPC algorithms followed by the EV CPO are presented.

\subsection{eMPC: EV CPO Cost Minimization}

This strategy, known as eMPC, seeks to minimize the operational costs of the CPO for the horizon $H$, by optimizing the EV charging profiles of the pool the interval $[k, ~k+H]$. The algorithm's decisions rely on the RT energy prices $\pi^c_k$ and $\pi^d_k$ for charging and discharging, respectively. Therefore, the cost function for V2G in eMPC is formulated as follows:
\begin{equation}
    \label{eq:CostMin}
    O^{\textit{CM}}_k = \pi^c_k \sum_{i=1}^{I} P^c_{i,k} - \pi^d_k \sum_{i=1}^{I} P^d_{i,k}
\end{equation}

Then, the EV CPO deals with the following eMPC V2G problem:
\begin{subequations} \label{eq:eMPC}
	\begin{eqnarray} \label{eq:eMPCfunc}
	\min_{P^c_{i,k}, P^d_{i,k}, z_{i,k}} & & \Delta t \sum_{k=0}^{H-1}  O^{\textit{CM}}_k  \\ \nonumber
 	\mbox{s.t.}  & & \text{Eq.~} \eqref{eq:EVSE}, \ \text{Eq.~} \eqref{eq:ConsSoC}, \ \text{Eq.~} \eqref{eq:SoC}, \ \text{Eq.~} \eqref{eq:ConsTra} \\ \nonumber
	& & 0 \leq P^c_{i,k} \leq \overline{P}^c_{i} z_{i,k} \\ \label{eq:eMPCconst}
	& & 0 \leq P^d_{i,k} \leq \overline{P}^d_{i} (1-z_{i,k})\\ \nonumber
	& & \forall \ k = 1,2,\dots,H, \ i = 1,2,\dots,I, \\ \nonumber
        & &  \ \ \ j = 1,2,\dots,J, \ g = 1,2,\dots,G.  \nonumber
	\end{eqnarray}
\end{subequations}

The decision variables of the problem include the charging ($P^c_{i,k}$) and discharging ($P^d_{i,k}$) power profiles, and the binary variables $z_{i,k}$, which determines whether charging or discharging is performed during time slot $k$. Both the charging and discharging powers are constrained to be between zero and their respective maximum powers, denoted as $\overline{P}^c_{i}$ and $\overline{P}^d_{i}$, respectively. Note that problem \eqref{eq:eMPCfunc} is Mixed-Integer Linear Program (MILP), for which efficient solvers are available. 

Notice that this eMPC problem also applies to the G2V strategy, with the only difference being the absence of the remuneration price $\pi^d_k$ since $P^d_{i,k}$ is always zero. Similarly, the binary variable $z_{i,k}$ is not required, resulting in a much simpler Linear Program (LP), with fewer decision variables.
 
\subsection{OCMF: CPO Flexibility Maximization}

This strategy denoted as OCMF, aims to achieve two objectives simultaneously. Firstly, it aims to minimize the operational cost of the charging pools, as presented in Eq.~\eqref{eq:CostMin}. Secondly, it aims to maximize the flexibility of the chargers' power.

The base definitions of the charger flexibility for G2V charging strategies were porposed in~\cite{DiazL2019} as follows:
\begin{itemize}
    \item Upward flexibility: This refers to the power increase a charger can deviate from the nominal dispatched power, reducing the charging time of the EV.
    \item Downward flexibility: This refers to the power reduction a charger can deviate from the nominal dispatched power. This flexibility can be utilized up to a later time slot when the EV needs to be charged to achieve the required SoC by its departure time.
\end{itemize}

\begin{figure}
    \centering
    \includegraphics[width=0.9\linewidth]{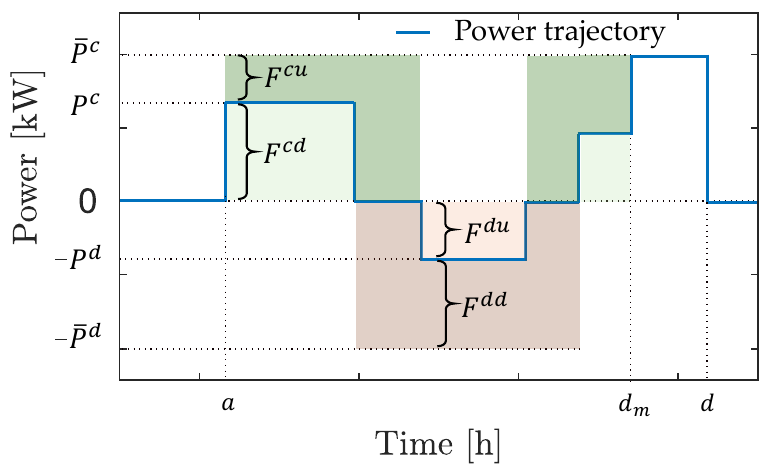}
    \caption{Flexibility in an EV charger, considering G2V and V2G strategies.}
    \label{fig:FlexV2G}
\end{figure}

Then, the G2V upward $F^{cu}_{i,k}$ and downward $F^{cd}_{i,k}$ flexibilities that each EV charger can provide are depicted in \figurename~\ref{fig:FlexV2G}. Then, we propose to define the charging flexibility $F^c_{i,k}$ as the minimum flexibility between upward and downward, as,
\begin{equation}
     F^c_{i,k} =  \textup{min} \{ F^{cu}_{i,k}, ~ F^{cd}_{i,k}\},
\end{equation}

In line with the charging flexibility discussed earlier, we introduce a discharging V2G flexibility $F^d_{i,k}$, designed to adjust the discharging power either upwards or downwards as needed by the CPO. Thus, to maximize both the upward $F^{du}_{i,k}$ and downward $F^{dd}_{i,k}$ discharging flexibilities, we consider the minimum of both as,
\begin{equation}
     F^d_{i,k} =  \textup{min} \{ F^{du}_{i,k}, ~ F^{dd}_{i,k}\},
\end{equation}

Note that only one of the charging or discharging flexibilities can be offered at each sample time when the EV is either charging or discharging. However, when the charger power is zero, both the charging and discharging flexibilities are available (refer to \figurename~\ref{fig:FlexV2G}). Furthermore, flexibility is not available after the time $d_m$ as the EV must be charged to reach the EV owner's requested SoC upon departure.

Therefore, the second objective of the OCMF, which aims to maximize the chargers' flexibility, is formulated as follows:
\begin{equation}
    O^{\textit{FM}}_k = \pi^c_{k,\textup{Flex}} \sum_{i=1}^{I} F^c_{i,k} + \pi^d_{k,\textup{Flex}} \sum_{i=1}^{I} F^d_{i,k},
\end{equation}
where, the $\pi^c_{k,\textup{Flex}}$ and $\pi^d_{k,\textup{Flex}}$ are the incentive prices for offering flexibility. 

The EV CPO addresses the following OCMF V2G problem:
\begin{subequations} \label{eq:OCCF}
    \begin{eqnarray} \label{eq:OCCFfunc}
    \min & & \Delta t \sum_{k=0}^{H-1}\left( O^{\textit{CM}}_k - O^{\textit{FM}}_k \right) \\ \nonumber
    \mbox{s.t.}  & & \text{Eq.~} \eqref{eq:EVSE}, \ \text{Eq.~} \eqref{eq:ConsSoC}, \ \text{Eq.~} \eqref{eq:SoC}, \ \text{Eq.~} \eqref{eq:ConsTra} \\  \nonumber
    & & F^c_{i,k} \leq P^c_{i,k} \leq \left( \overline{P}^c_{i} - F^c_{i,k} \right) z_{i,k}\\ \nonumber
    & & F^d_{i,k} \leq P^d_{i,k} \leq \left( \overline{P}^d_{i} - F^d_{i,k} \right) (1-z_{i,k}) \\ \label{eq:OCCFconst}
     & & 0 \leq F^c_{i,k} \leq \overline{P}^c_{i} \\ \nonumber
     & & 0 \leq F^d_{i,k} \leq \overline{P}^d_{i} \\ \nonumber
    & & \forall \ k {=} 1,2,\dots,H, \ i {=} 1,2,\dots,I, \\ \nonumber
        & &  \ \ \ j {=} 1,2,\dots,J, \ g {=} 1,2,\dots,G.  \nonumber
    \end{eqnarray}
\end{subequations}

The decision variables of this problem are the charging $P^c_{i,k}$ and discharging $P^d_{i,k}$ power, the binary variable $z_{i,k}$, and the charging $F^c_{i,k}$ and discharging $F^d_{i,k}$ flexibilities. The charging and discharging powers are constrained by the charger flexibility, with the charging and discharging flexibilities restricted to positive values and bounded between zero and the maximum power.

Notice that as in the eMPC problem, this OCMF problem also applies to the G2V strategy, with the difference being the absence of the remuneration price $\pi^d_k$ since $P^d_{i,k}$ is always zero, the lack of the price $\pi^d_{k,\textup{Flex}}$, the flexibility $F^d_{i,k}$, and the binary variable $z_{i,k}$.


\section{Experimental Setup}

\begin{algorithm}[t]
\caption{EV2Gym Simulation with eMPC V2G}
\label{alg:simulation}
\begin{algorithmic}[1]
\State \textbf{from} EV2Gym.ev2gym\_env \textbf{import} EV2Gym
\State \textbf{from} ocmf\_mpc \textbf{import} OCMF\_V2G, OCMF\_G2V
\State \textbf{from} eMPC \textbf{import} eMPC\_V2G, eMPC\_G2V
\State config\_file = ``config.yaml'' \Comment{Configuration file path}
\State env = EV2Gym(config\_file) \Comment{Initialize simulation}
\State $\boldsymbol{s}_0$ = env.reset()
\State agent = eMPC\_V2G(env, control\_horizon=$H$) 
\For{$k$ \textbf{in} 0, 1, \dots, $T$ }    
    \State $\boldsymbol{a}_{1,\dots,I}$ = agent.get\_action($\boldsymbol{s}_k$)    \Comment{MPC optimization}
    \State $\boldsymbol{s}_{k+1}$, stats = env.step($\boldsymbol{a}_{1,\dots,I}$) \Comment{Simulate}
\EndFor
\State profits, total energy charged, ... = stats
\end{algorithmic}
\end{algorithm}

In this Section we describe how to integrate the execution of an EV2Gym simulation, introduced in Section~\ref{sec:ev2gym}, with the proposed MPC algorithms. Alg.~\ref{alg:simulation} outlines the Python code necessary to run a simulation utilizing the \verb|eMPC_V2G| strategy. Initially, the simulator class \verb|EV2Gym| (line $1$) and the developed MPC classes \verb|OCMF_V2G|, \verb|OCMF_G2V|, \verb|eMPC_V2G|, and \verb|eMPC_G2V| (lines $2$ and $3$) are imported. Lines $5$ and $6$ initialize the simulation environment using a user-defined configuration file (\verb|config_file|). Subsequently, the MPC algorithm's variables are set up in line $7$, allowing users to customize the control horizon $H$ as needed.

The main simulation loop, spanning $T$ steps, commences thereafter. Firstly, charging and discharging actions $\boldsymbol{a}_{1,\dots,I}$ for each charging station $i \in I$ are determined by executing the MPC algorithm's optimization process (line $9$), using the current state ($\boldsymbol{s}_k$) of the simulation as input. The state $\boldsymbol{s}_k$ includes essential information such as EVs' SoC, time of arrival and departure, load and PV forecasts, facilitating the decision-making algorithm's optimization. The actions vector $\boldsymbol{a}{1,\dots,I}$ is constrained within the $[-1, 1]^I$ interval, where $-1$ denotes maximum power discharge, $1$ signifies maximum power charging, and $0$ indicates no action. Subsequently, a simulation step is taken (\verb|env.step()|) utilizing the action vector $\mathbf{a}{1,\dots,I}$ as a parameter (line 10). Each step yields the subsequent simulation state $\boldsymbol{s}_{k+1}$, while the final step ($k=T$) also provides simulation evaluation statistics (\verb|stats|) such as total profits (€), total energy charged and discharged (kWh), execution time (s), and total battery degradation (line 12). These metrics are important for evaluating the efficacy of smart charging algorithms.

\subsection{Experimental Configuration}

The EV2Gym simulator, allows us to fully configure the dynamics of the simulation by modifying the configuration file. Table~\ref{tab:SimPar} showcases the general simulation parameters and the values we used for our experiments. Specifically, our simulations last 24 hours, with discrete time steps set at 15 minutes. Each simulation contains the connection of 5 to 60 EVSEs to either $1$ or $3$ transformers, accommodating a fleet size ranging from 15 to 120 EVs, as determined by our experimental setups. The EVs' time of arrival, time of departure, and arrival SoC are based on residential EV transaction data from ElaadNL~\cite{elaad}. Also, the discharge price is based on the charge price and multiplier $m$, thus $\pi^d_k = m \cdot \pi^c_k$. The charging prices used in these simulations are based on the historic Dutch day-ahead prices~\cite{entsoe}.

\begin{table}[t]
    \centering
    \small
    \caption{General Simulation parameters.}
    \label{tab:SimPar}
    \begin{tabular}{lcc}
	\hline
	\textbf{Name} & \textbf{Symbol} & \textbf{Value} \\
	\hline
        Sample time [min] & $\Delta t$ & 15  \\
        Operation time of the station [h] & $T$ & 24 \\
        Prediction horizon (2.5~h - 10~h) & $H$ & \{2.5$\times$4, 10$\times$4\} \\
        Number of EVSEs & $I$ & \{5 - 50\} \\
        Number of EVs & $J$ & \{15 - 120\} \\
        Number of Transformers & $G$ & \{1, 3\}\\        
        Discharge Price multiplier & $m$ & \{0.8 - 1.2\}\\
        EV Scenario &  & ``Residential''\\
	\hline
    \end{tabular}
\end{table}

Furthermore, the default model parameters, such as transformer power limit, EVSE, and EV characteristics, can be found in Table~\ref{tab:model_config}. Finally, the transformer loads, generation, and demand response parameters can be found in Table~\ref{tab:tr_config_parameters}. The normalization multipliers refer to how many times greater the inflexible load or generation peak (kW) is compared to the transformer power limit. For example, $1$ means that the maximum inflexible load will be as much as the transformer limit. The simulator also provides forecasts of the loads using a Gaussian distribution $(\bar{l}_{g,k} = \mathcal{N}(\mu,\sigma))$ with mean the $\mu = 100\% \times {l}_{g,k}$ of the actual load at time $k$ and standard deviation $\sigma = 5\% \times {l}_{g,k}$ of it. Similarly, for PV generation forecasts $\bar{w}_{g,k}$. Moreover, tool users have the flexibility to customize demand response events by specifying parameters such as the number of events, capacity reduction, duration, and notification time. The start time of events follows a Gaussian distribution with a mean of 18:00 and a standard deviation of 1 hour.

\begin{table}[t]
    \centering
    \small
    \caption{Default Simulation Model Parameters.}
    \label{tab:model_config}
    \begin{tabular}{lcc}
	\hline
	\textbf{Name} & \textbf{Symbol} & \textbf{Value} \\
	\hline
        Transformer power limit [kW] & $\Psi$ & $400$  \\        
        Maximum EVSE output power [kW] & $ \overline{P}^c$, $ \overline{P}^d$ &  $22$\\ 
        EVSE Voltage (V) & $V$ & $230$ \\
        EVSE Phases & $\phi$ & $3$  \\        
        EV Battery Capacity [kWh] & $E_j$ & $50$\\       
        Maximum EV power [kW] & $\overline{P}^c$, $ \overline{P}^d$  &  $22$\\ 
        Minimum EV SoC when discharging & $SoC_{j,min}$ & $10\%$\\ 
        Minimum EV SoC at departure & $ \underline{SoC}_{j, d_j}$ & $80\%$\\
        Minimum EV time of connection [h] & & $3$\\
        Charging Efficiency & $\eta^c$  &  $100\%$\\ 
        Discharging Efficiency & $\eta^d$  &  $100\%$\\  
	\hline
    \end{tabular}
\end{table}

\begin{table}[t]
    \centering
    \small
    \caption{Default Transformer Loads Parameters.}
    \label{tab:tr_config_parameters}
    \begin{tabular}{lcc}
	\hline
	\textbf{Name} & \textbf{Value} \\
	\hline
        Inflexible Loads normalization multiplier & $1$  \\         
        Inflexible Loads forecast mean & $100\%$  \\ 
        Inflexible Loads forecast standard deviation & $5\%$  \\ 
        PV Generation normalization multiplier & $3$  \\         
        PV Generation forecast mean & $100\%$  \\ 
        PV Generation forecast standard deviation & $5\%$  \\ 
        Demand Response duration [h] & $1$ \\
        Demand Response in advance notification [min] & $15$ \\
        Number of Demand Response events in a day & $1$ \\
        Demand Response capacity reduction & $20\%$ \\
        Demand Response event start mean [h] & $18$ \\
        Demand Response event start standard deviation [h] & $1$ \\        
	\hline
    \end{tabular}
\end{table}

\section{Experimental Results}
\label{sec:results}

In this section, we provide a thorough experimental assessment of the proposed MPC strategies. Initially, we delve into a detailed examination of a single case study to understand the behaviors. Subsequently, we broaden our scope by conducting multiple simulations to show average performance, facilitating a comparative analysis across the main evaluation metrics including profitability, flexibility, and battery degradation. Finally, we offer an evaluation of the computational costs associated with the proposed methods, thereby providing a holistic perspective on their efficacy.

\begin{figure*}[t]
     \centering
     \subfloat[Charging and discharging prices.]{
         \centering
         \includegraphics[width=0.671\linewidth]{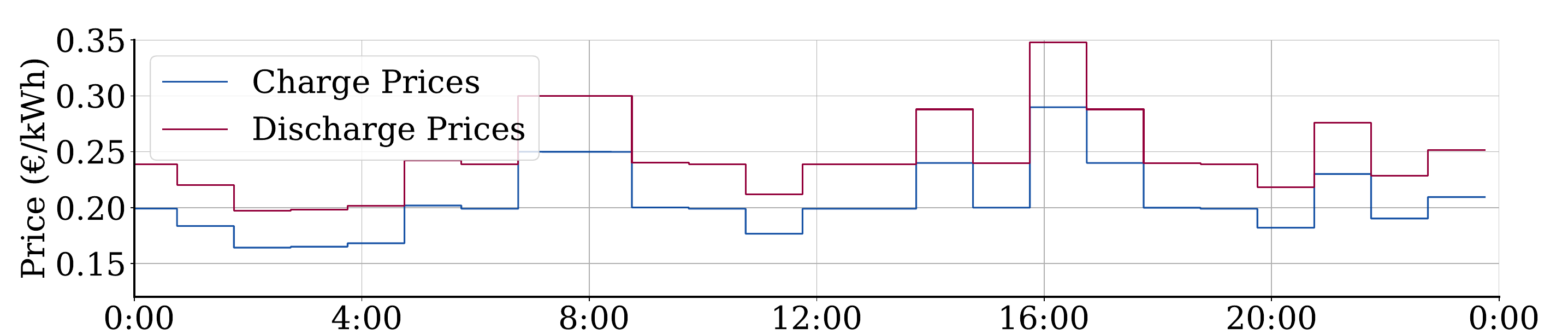}
         \label{fig:prices}
         }
         \\
         \subfloat[SoC of EVs in a single EVSE.]{
         \centering
         \includegraphics[width=0.67\linewidth]{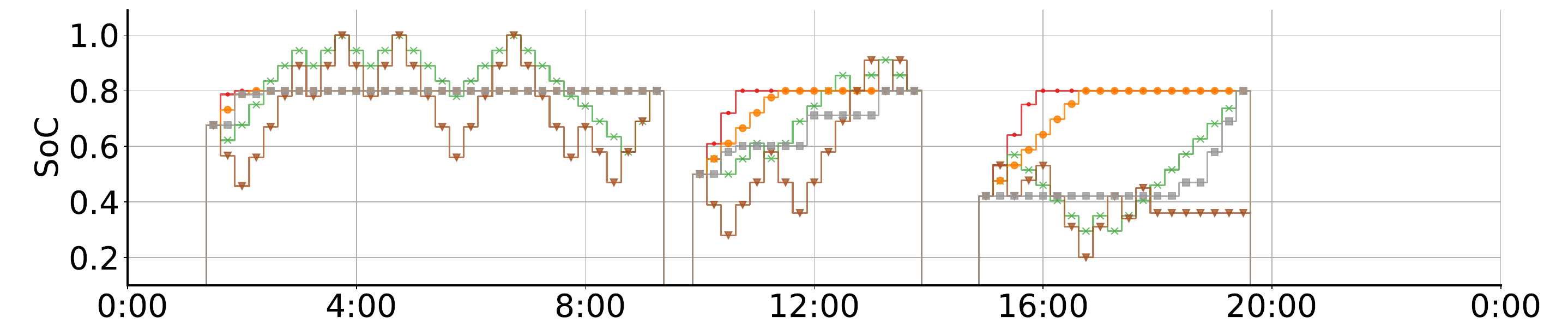}
         \label{fig:ev_soc}
         }
                  \\
         \subfloat[Power output the same EVSE.]{
         \centering
         \includegraphics[width=0.67\linewidth]{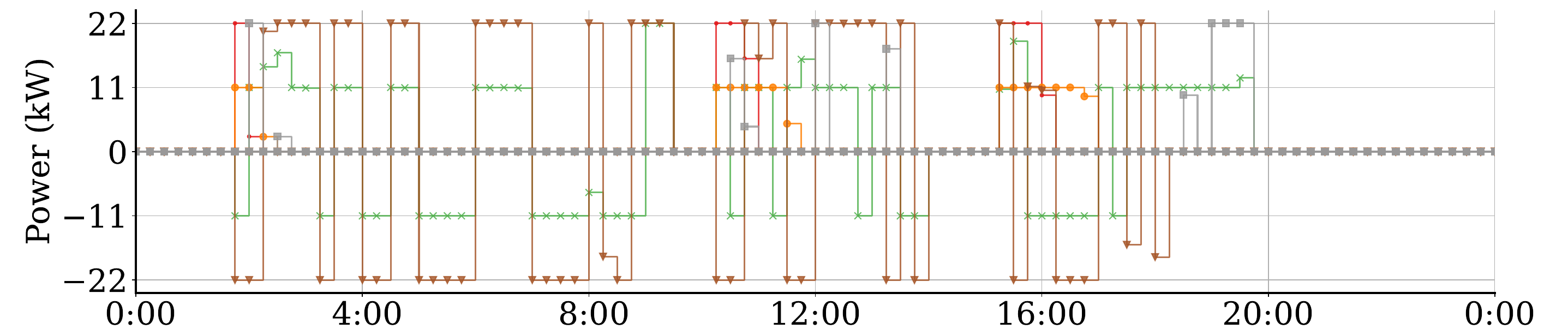}
         \label{fig:cs_power}
         }
         \\
     \subfloat[Total power comparison for 30 EVSEs.]{
         \centering
         \includegraphics[width=0.67\linewidth]{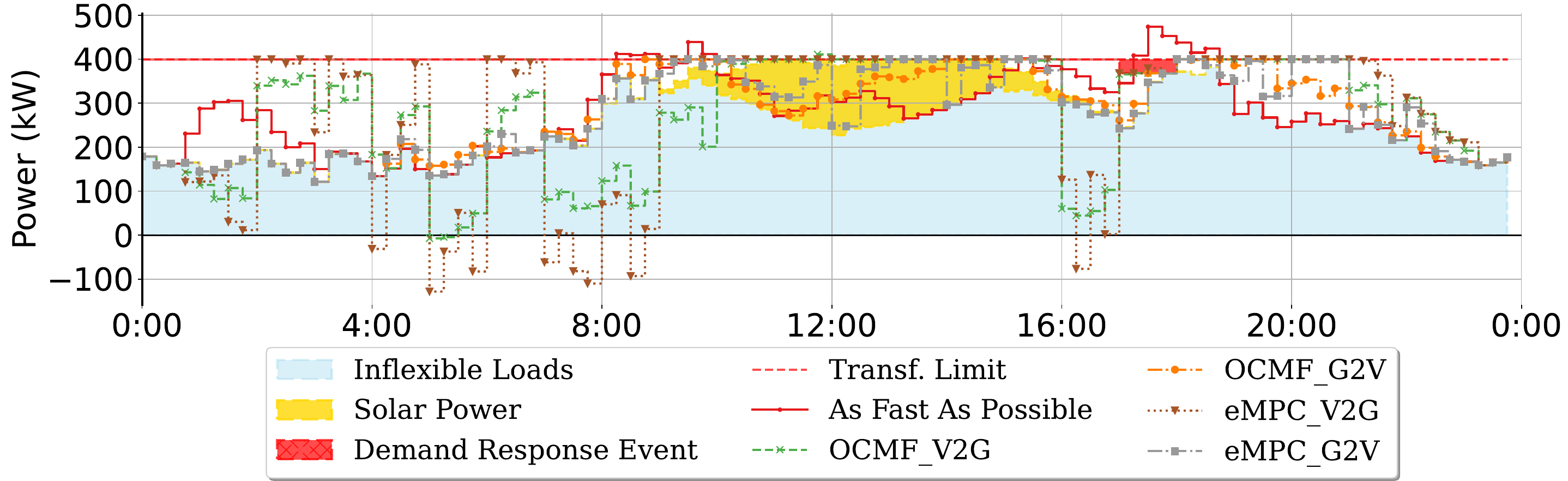}
         \label{fig:power_comparison}}
        \caption{Comparison of proposed MPC approaches for a case study with $30$ EVSEs connected to a single transformer, and  $75$ EVs.
        }
        \label{fig:v2gwithloads_comp_}
\end{figure*}
\subsection{Performance Comparison}

To begin our assessment of the proposed MPC methods, we first compare their performance against baseline strategies in a single case study. In Fig.\ref{fig:v2gwithloads_comp_}, we present a comparison between the \textit{charge as fast as possible} strategy (AFAP) and the proposed MPC methods with a prediction horizon of 10 hours ($H=40$), examining four key aspects: pricing dynamics, SoC of EVs, charging station power output, and aggregated power at the transformer level. Fig.\ref{fig:prices} displays the charging and discharging prices observed during the simulation day, with slightly higher prices for discharging to incentivize V2G operations. Additionally, Fig.\ref{fig:ev_soc} illustrates the fluctuation in SoC of EVs connected to an EVSE throughout the day, highlighting the strategies employed to maximize profits (eMPC) or flexibility (OCMF). Fig.\ref{fig:cs_power} shows the power output profile of the same EVSE, indicating that eMPC methods utilize the maximum power output ($22$ kWh), whereas OCMF typically operates around $11$ kWh. This behavior reflects the MPC methods' focus on maximizing flexibility.

Furthermore, in Fig.~\ref{fig:power_comparison}, we present a comparison of the total power output from all methods controlling 75 EVs that are connected to $30$ EVSEs over a day. This comparison accounts for inflexible loads, PV power generation, and a demand response event. It is evident how the AFAP strategy risks transformer overloads, whereas the MPC methods ensure smooth operation by effectively utilizing the transformer's capacity. Additionally, the MPC approaches demonstrate adaptability by reducing power usage to adhere to a demand response event occurring at 18:00. Notably, the demand response event is known only 15 minutes (1 step) in advance, underscoring the necessity for MPC's real-time control capabilities.


  
\subsection{Performance Analysis}

In the previous section, we evaluated the performance of the proposed approaches within a single case study. However, a comprehensive assessment is necessary across various scenarios. Table~\ref{tab:sensitivity_analysis} presents the mean and standard deviation derived from $50$ statistical runs across a configuration comprising $10$ bidirectional chargers linked to one transformer, around $25$ EVs per simulation, and discharge price multiplier $m=1.2$. These case studies included elements such as inflexible loads, PV generation, and demand response events, thereby mirroring real-world optimization challenges. Across all scenarios, the MPC methods employed a 10-step prediction horizon (i.e., 2.5~h, $H=10$) to drive their decision-making processes.

\begin{table}[t]
\centering
\caption{Averages of $50$ runs with 10 EVSEs, 25 EVs, 1 transformer, and varying EV behavior.}
\label{tab:sensitivity_analysis}
\resizebox{\linewidth}{!}{%
\begin{tabular}{ccccccccc}
\hline
Algorithm & \begin{tabular}[c]{@{}c@{}}Profits\\ (\euro)\end{tabular} & \begin{tabular}[c]{@{}c@{}}Energy\\  Charged/ Discharged \\ (kWh)\end{tabular}
& \begin{tabular}[c]{@{}c@{}} $\sum Q^\text{lost}$ \\ ($\times10^{-4}$)\end{tabular} &\begin{tabular}[c]{@{}c@{}} $\sum d^\textit{cal}$ \\ ($\times10^{-4}$)\end{tabular} &\begin{tabular}[c]{@{}c@{}} $\sum d^\textit{cyc}$ \\ ($\times10^{-4}$) \end{tabular}\\ \hline
AFAP & $-47.6$ ±$8.4$ & $227$ ±$38$ / - & $12.4$ ±$1.6$ & $3.5$ ±$0.2$ & $8.8$ ±$1.5$ \\
OCMF G2V & $-48.0$ ±$8.3$ & $227$ ±$38$ / - & $12.8$ ±$1.6$ & $3.3$ ±$0.2$ & $9.5$ ±$1.5$ \\
OCMF V2G & $11.9$ ±$8.8$ & $927$ ±$73$ / $706$ ±$63$ & $36.8$ ±$2.7$ & $3.1$ ±$0.2$ & $33.7$ ±$2.5$ \\
eMPC G2V & $-43.9$ ±$7.6$ & $227$ ±$38$ / - & $13.3$ ±$1.7$ & $3.3$ ±$0.2$ & $10.0$ ±$1.6$ \\
eMPC V2G & $50.8$ ±$10.3$ & $1390$ ±$119$ / $1172$ ±$107$ & $ 57.1 $ ±$4.8$ & $3.2$ ±$0.2$ & $53.9$ ±$4.5$ \\ \hline
\end{tabular}%
}
\end{table}


Table~\ref{tab:sensitivity_analysis} illustrates that across all G2V approaches (AFAP, OCMF G2V, and eMPC G2V), the average energy consumption remains consistent at $227$ ±$38$ kWh. However, OCMF G2V stands out for its ability to consistently maximize flexibility, ensuring swift responses to demand response events. On the other hand, eMPC G2V excels in cost optimization, reducing charging expenses from $48$ to $43.9$\euro. While the direct cost savings of eMPC G2V compared to OCMF G2V are only around $4$\euro, the significance lies in OCMF G2V's capacity to provide flexibility, underscoring its critical importance. Meanwhile, when V2G is utilized both OCMF V2G and eMPC V2G, end up with a profit of around $11.9$ and $50.8$\euro, respectively. This shows that when V2G is utilized it is possible to support the grid and make significant profits.
\begin{figure}[!t]
         \centering
         \includegraphics[width=1\linewidth]{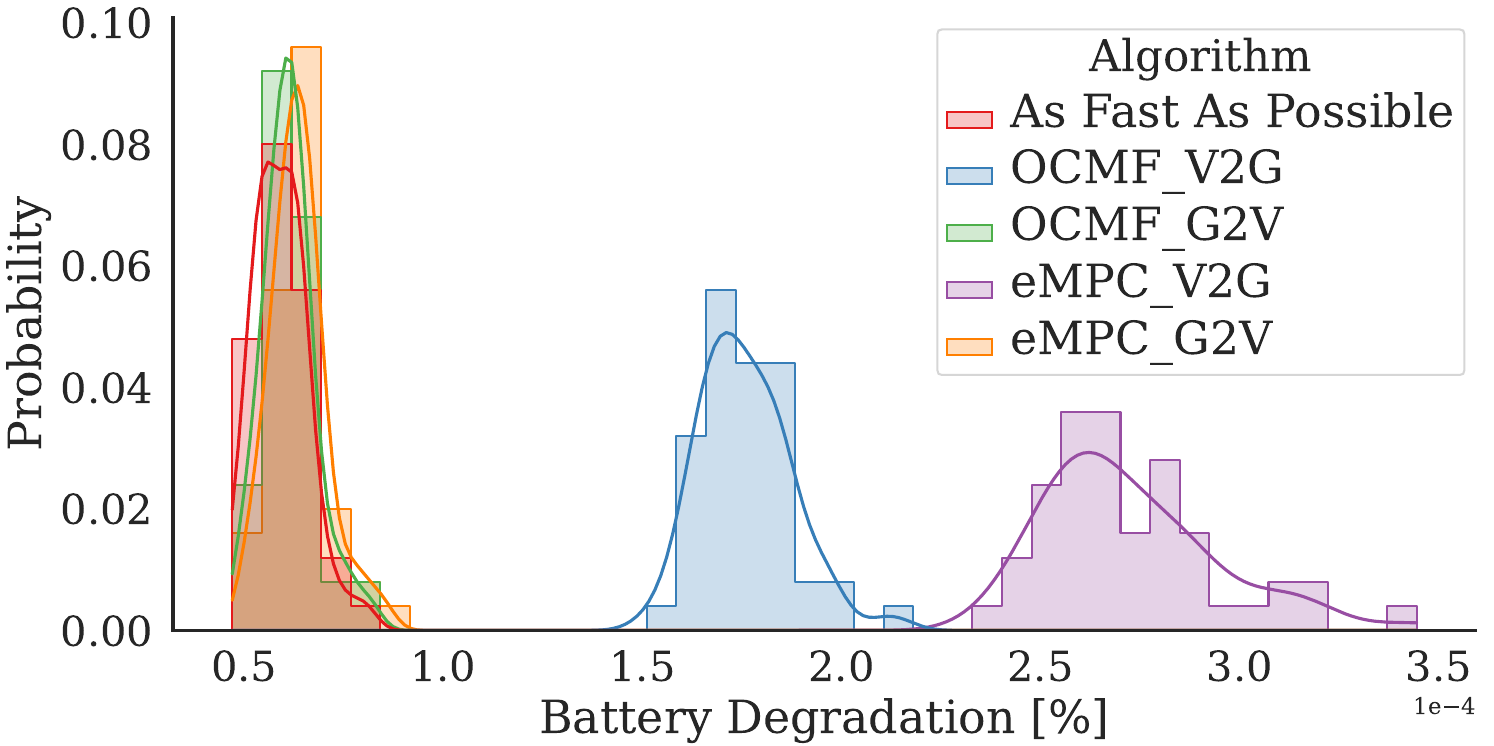}
        \caption{Average capacity loss per EV because of battery degradation.
        }
        \label{fig:battery_degradation}
\end{figure}

Moreover, it is crucial to evaluate the impact of these methods on total battery degradation. Table~\ref{tab:sensitivity_analysis} provides insights into the total battery capacity lost ($Q^\textit{lost}$) during the simulation for all EVs, split into calendar aging ($d^\textit{cal}$) and cyclic degradation ($d^\textit{cyc}$) components, while Fig.~\ref{fig:battery_degradation} depicts histograms of the percentage ($10^{-4} \times 100\%$) average capacity lost per EV.
Remarkably, V2G methods tend to mitigate calendar aging due to their maintenance of a lower average SoC during the charging phase. However, this advantage is counterbalanced by substantially higher cyclic degradation values (up to $45 \times 10^{-4}$ higher), stemming from their utilization of EV batteries for both charging and discharging, totaling $2562$ kWh—$2335$ kWh more than G2V methods. Conversely, all G2V methods display comparable battery degradation ($\approx 13 \times 10^{-4}$), with OCMF G2V offering maximal flexibility and eMPC G2V focusing on minimizing charging costs. Overall, these findings underscore the capabilities of the proposed algorithms and their intricate trade-offs among profits, flexibility, and battery degradation.

\subsection{Cost Sensitivity Analysis}

\begin{figure}[!t]
         \centering
         \includegraphics[width=1\linewidth]{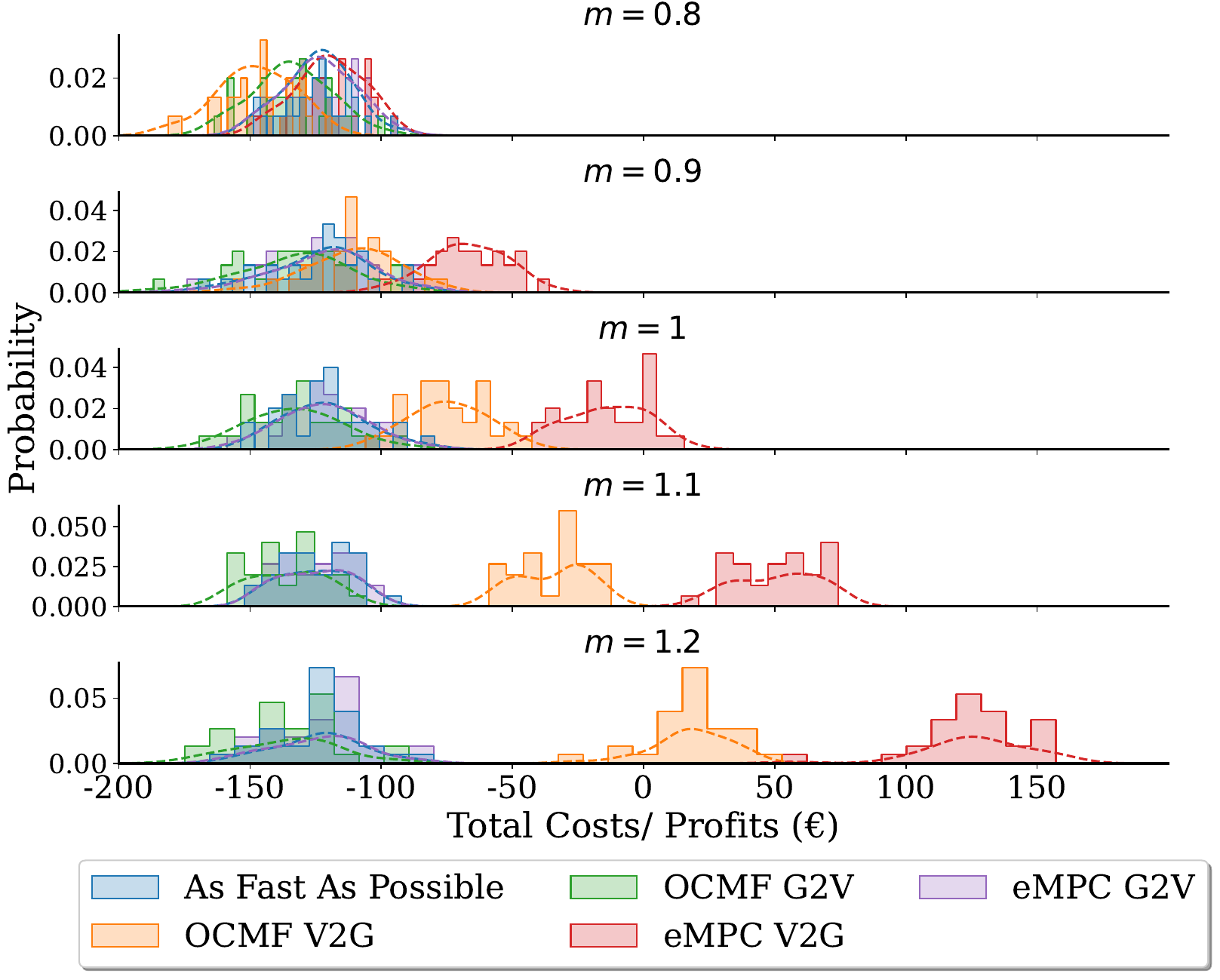}
        \caption{Comparison of EV-user charging costs across different discharge price multipliers ($m$). Note that these figures solely reflect EV-user payments. However, it's important to acknowledge that potential profits from flexibility provision are not included here, which could significantly boost overall profits.}
        \label{fig:cost_analysis}
\end{figure}

In the previous experiments, we observed how the V2G methods can be profitable when the discharge price multiplier is set to $m=1.2$; however, it is important to conduct a sensitivity analysis on alternative price scenarios. In each scenario, the price is defined as:
\begin{equation}
 \pi^d_k = m \cdot \pi^c_k,
 \end{equation}
for $k=1,\dots,T$. Fig.\ref{fig:cost_analysis} illustrates the probability density functions of the total costs and profits for $50$ statistical runs for each price scenario. In every scenario, the G2V and AFAP methods have the same distribution since they are not affected by the discharging prices. As observed in Fig.\ref{fig:cost_analysis}, the eMPC\_V2G strategy always provides the highest cost reductions ($m={0.8,0.9}$) and the highest profits ($m={1, 1.1, 1.2}$). OCMF\_V2G is also profitable in all cases except $m=0.8$, showcasing on average around $10$\euro~higher charging costs; this happens because the cost of providing flexibility was less than discharging, hence leading to this behavior. Notice that in this cost/profit analysis of the OCMF strategy, the payments for flexibility have not been included. Therefore, it can be inferred that this strategy would significantly increase its profits when the flexibility offered and provided is considered.

\subsection{Computational Cost}

\begin{figure}[t]
     \centering
     \subfloat[$h=10$]{
         \centering
         \includegraphics[width=0.49\linewidth]{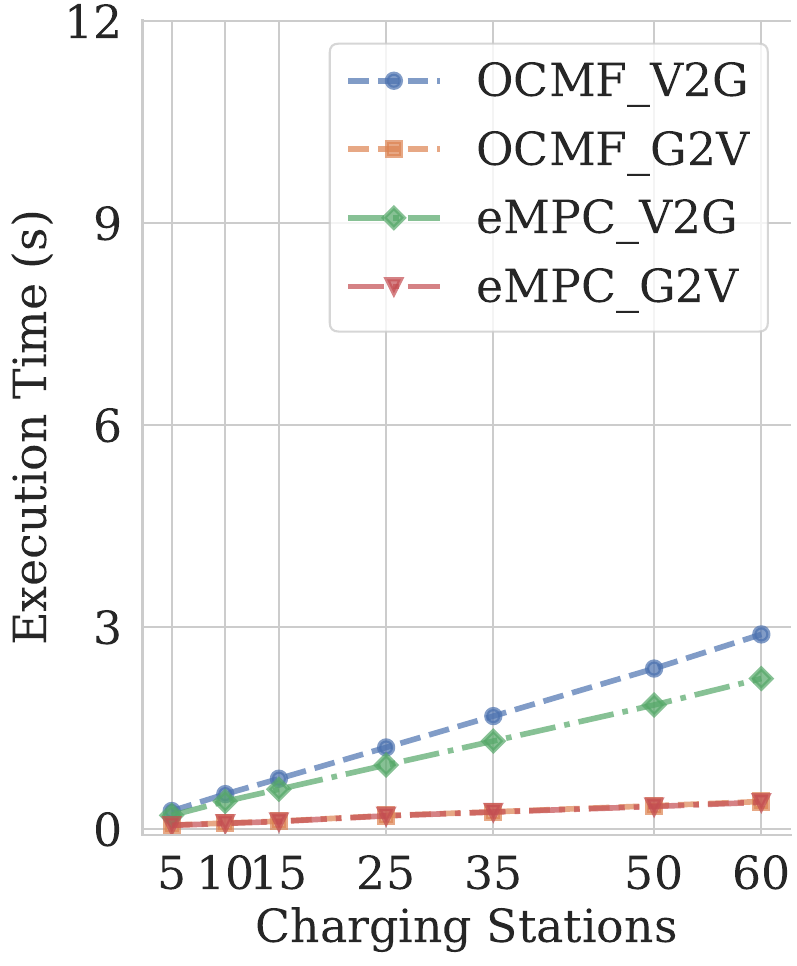}
         \label{fig:exec_time_10}
         }
         \subfloat[$h=30$]{
         \centering
         \includegraphics[width=0.48\linewidth]{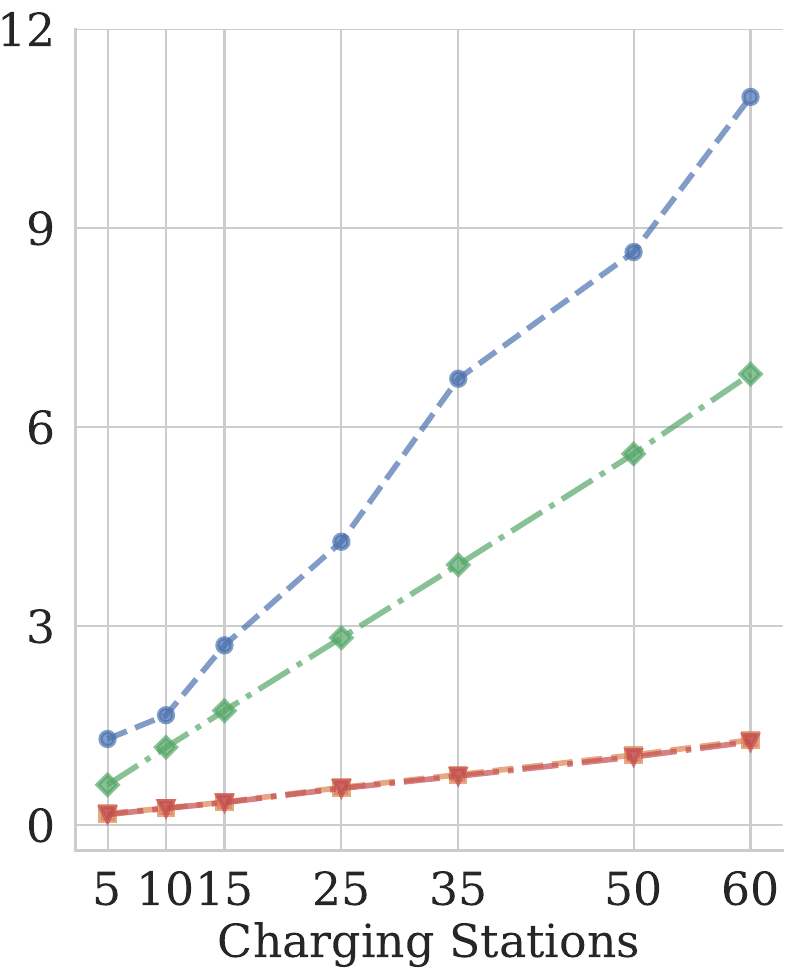}
         \label{fig:exec_time_30}
         }
        \caption{Average step execution time for varying control horizons and EVSEs.
        }
        \label{fig:exec_time}
\end{figure}

The proposed MPC methods tackle real-time EV smart-charging challenges efficiently. However, it's noteworthy that as the number of decision variables and constraints increases, the execution time of these approaches increases too. To evaluate the performance of each method, we conducted an assessment by measuring their execution times using high-performance hardware setups. Specifically, we utilized the ``AMD RYZEN 7 5700X 8-Core'' processor paired with 32GB of RAM, and employed the Gurobi solver.
Fig.~\ref{fig:exec_time} illustrates the comparison of average per-step execution times for various case studies, with EVSEs ranging from 5 to 60, all connected to three transformers. The execution time mentioned includes the time needed for creating the matrices, creating the model with the constraints, and then solving it having the MIP gap~\footnote{Notice that, by using a higher MIP Gap can significantly reduce the execution time, but at the cost of a solution quality.} parameter set to 0.
Initially, we can observe in Fig.~\ref{fig:exec_time_10}, the slowly increasing step time when using 10-step prediction horizons. On the other hand, in Fig.~\ref{fig:exec_time_30}, we see a larger growth in per-step execution time with the 30-step prediction horizon.
As expected, OCMF V2G showcases the largest increase in execution time in both cases as it has the most decision variables compared to the other approaches.
Notably, in both scenarios, the execution time for G2V methods scales at a slower rate, requiring only half the time compared to V2G methods. Moreover, notice that in the worst-case scenario, the execution time of the MPC algorithm is less than $1.5\%$ of the time step of the control strategy, leaving room to manage larger pools of chargers in real time. 

\section{Conclusions}
In this paper, we introduced four variations of MPC tailored to address real-time EV G2V and V2G smart charging, while accounting for transformer loads and demand response events. Through a comprehensive evaluation across diverse case studies, we examine the effectiveness of our methods. Furthermore, we analyze the impact of these strategies on EV battery degradation, highlighting the intricate balance between profitability, operational flexibility, and battery degradation.

\bibliographystyle{IEEEtran}
\bibliography{ref}

\end{document}